\documentclass[usenatbib]{mnras}
\usepackage{graphicx}
\usepackage{float}
\usepackage{epsfig}
\usepackage{times}
\usepackage{longtable}
\usepackage{afterpage}
\usepackage{pdflscape}
\usepackage{amsmath}
\usepackage{amssymb}
\usepackage{comment}

\begin{document}

\title{The sub-galactic and nuclear main sequences for local star-forming galaxies}

\author[A. Maragkoudakis et al.]
{A. Maragkoudakis,$^1$ $^2$ $^3$ A. Zezas,$^1$ $^2$ $^3$ M. L. N. Ashby,$^2$ S. P. Willner$^2$ \\
$^1$University of Crete, Department of Physics, Heraklion 71003, Greece \\
$^2$Harvard-Smithsonian Center for Astrophysics, Cambridge, MA 02138 \\
$^3$Foundation for Research and Technology - Hellas (FORTH), Heraklion 71003, Greece}

\maketitle
\begin{abstract}
We describe a sub-galactic main sequence (SGMS) relating star formation rate surface density ($\Sigma_{\textrm{SFR}}$) and stellar mass density ($\Sigma_{\star}$) for distinct regions within star forming galaxies, including their nuclei. We use a sample of 246 nearby star-forming galaxies from the ``Star Formation Reference Survey" and  demonstrate that the SGMS holds down to $ \sim $1 kpc scales with a slope of $\alpha=0.91$ and a dispersion of 0.31\,dex, similar to the well-known main sequence (MS) measured for globally integrated star formation rates (SFRs) and stellar masses. The SGMS slope depends on galaxy morphology, with late-type galaxies (Sc--Irr) having $\alpha = 0.97$ and early-type spirals (Sa--Sbc) having $\alpha = 0.81$. The SGMS constructed from sub-regions of individual galaxies has on average the same characteristics as the composite SGMS from all galaxies. The SGMS for galaxy nuclei shows a dispersion similar to that seen for other sub-regions. Sampling a limited range of SFR--M$ _{\star} $ space may produce either sub-linearity or super-linearity of the SGMS slope. For nearly all galaxies, both SFR and stellar mass peak in the nucleus, indicating that circumnuclear clusters are among the most actively star-forming regions in the galaxy and the most massive. The nuclear SFR also correlates with total galaxy mass, forming a distinct sequence from the standard MS of star-formation. The nuclear main sequence will be useful for studying bulge growth and for characterizing feedback processes connecting AGN and star formation.

\end{abstract}


\section{INTRODUCTION}

The tight relationship between galaxy star-formation rate (SFR) and total stellar mass ($M_{\star}$), commonly referred to as the star-forming galaxy main sequence (MS), has been extensively investigated over the past decade (e.g., \citealt{Noeske07}, \citealt{Elbaz07}, \citealt{Daddi07}). The MS describes the ratio of current to past star formation and indicates that higher stellar-mass systems undergo more intense star formation activity than lower-mass systems. The low dispersion of the MS suggests that star formation is regulated mostly through secular processes rather than stochastic merger-driven star-forming episodes. A general description of the MS is a power law of the form SFR $\propto M_{\star}^{\alpha}$ with typical values of $ \alpha $ between 0.6 and 1 (\citealt{Rodighiero11}). 

The determination of the MS slope depends strongly on the prescriptions used to measure SFR and $M_{\star}$, luminosity to SFR conversions and stellar population synthesis (SPS) models. This is true even for samples in narrow redshift ranges, where significant variations (up to $\sim$0.35~dex) are reported between MS slopes derived from different studies (e.g., \citealt{Speagle14} (S14)). Additional complications arise from the selection criteria for the star-forming galaxy (SFG) samples. For example using the \textit{BzK} two-color selection (\citealt{Daddi07}), or selecting blue cloud galaxies by making a cut in the color-magnitude diagram (CMD) (\citealt{Elbaz07}), is biased toward more actively star-forming systems missing the non-negligible fraction of galaxies at the lower end of the star-formation range.

More recently, a few studies (e.g., \citealt{Wuyts13} (W13); \citealt{Hemmati14}; \citealt{Magdis16} (M16); \citealt{CanoDiaz16} (C16)) provide evidence that MS-like correlations are also present at sub-galactic scales. These studies compared SFR surface densities with stellar-mass surface densities for individual sub-galactic regions. While at intermediate to higher redshifts ($z > 1$) this correlation has a slope close to unity (W13, M16), at lower redshifts sub-linearity has been reported (C16). \cite{Delgado16} (D16) showed that local specific SFR (sSFR) shows radial profiles that increase outwards and scale with Hubble type. Specifically, the sSFR in the inner 1 half light radius is steeper than outwards, suggesting that galaxies are quenched inside-out and that this process is faster in the central part, causing the observed sub-linearity in the MS slope. However, existing studies concentrated mostly on the higher end of the galaxy stellar mass distribution, or they covered a small range in SFR and/or stellar mass. In addition, these studies relied on spectral energy distribution (SED) fitting or stellar population models for the determination of SFR and stellar masses, the results of which depend on parameters such as the choice of star formation history (SFH) priors or the parameters of the stellar templates used (e.g., \citealt{Conroy13}). In addition, these methods suffer from degeneracies where several combinations of different models are able to reproduce the observables, and therefore the derived galaxy parameters can be highly uncertain (e.g., \citealt{Conroy10}).


This work explores star formation taking place in the sub-galactic regions of a representative sample of 246 SFGs in the local Universe ($z < 0.08$) with respect to their local and total stellar mass content. The sample spans a range of four orders of magnitude in both stellar mass and SFR, greatly expanding the parameter space covered by previous studies. In addition, we explore the MS in the nuclear/circumnuclear region in the context of the general sub-galactic MS (SGMS). Throughout the paper, SGMS will be used to refer to the sequence of local SFR density and local stellar-mass density formed from all sub-galactic regions in galaxies including nuclei, nuclear SGMS to the corresponding sequence formed from nuclear regions only, and nuclear MS (NMS) to the sequence of nuclear SFR and total stellar mass. 

This paper is organized as follows: Section \ref{S-A} describes the sample used in our analysis and the methods used to derive SFRs and stellar masses. Section \ref{globalMS} describes the integrated MS of our SFG sample and demonstrates the validity of the SFR and $M_{\star} $ measures used. Section \ref{SGMS} presents the SGMS and its dependence on galaxy morphology (Section \ref{morphology}) and the nuclear/circumnuclear SGMS (Section \ref{nucleus}). Section \ref{Discussion} discusses our results in the context of local and higher-$z$ galaxies. Throughout this paper we adopt a $ \Lambda $CDM cosmology with $H_{0} = 73$ km s$^{-1}$, $\Omega_{\textrm{M}} = 0.3$, $\Omega_{\Lambda} = 0.7$, and a \cite{Salpeter55} initial mass function (IMF) for SFRs and stellar mass determination.

\section{Sample Description and Methodology} \label{S-A}

\subsection{The SFRS sample} \label{Sample}

The Star Formation Reference Survey (SFRS; \citealt{SFRS}) is representative of the full range of star-formation properties and conditions in the nearby Universe. It consists of 369 galaxies selected from the PSCz catalog (\citealt{Saunders00}) in order to sample the entire range of SFR, specific SFR (sSFR), and interstellar dust temperature as represented by the 60 $\mu$m luminosity, the flux ratio of $F_{60}/K_{S}$, and the far-IR flux density ratio $F_{100}/F_{60}$ respectively.

Owing to the representative nature of the SFRS sample, it includes bona-fide SFGs and AGNs. The latter were retained in order not to bias the sample. Therefore, in order to identify star forming galaxies, Maragkoudakis et al. (in preparation) used a combination of IR diagnostics (\citealt{Stern05}) and optical emission-line diagnostics (i.e., BPT diagrams; \citealt{BPT}, \citealt{Kewley01}, \citealt{Kauffmann03}) while accounting for the uncertainties of the prominent emission-lines used in the optical diagnostics. The results of this analysis showed that the SFRS sample consists of 260 SFGs, 49 Seyfert, 34 composite objects, and 26 LINERs. In this work we use 246 SFRS SFGs with $ z < 0.08 $ and available data to probe the stellar mass and SFR content in global and spatially resolved sub-galactic scales. The remaining 14 galaxies lacked data in one or more IRAC bands, or have unreliable SDSS color information, or are too distant to meet the resolution thresholds (see Section \ref{Analysis}) applied for the sub-galactic analysis. The final sample covers the redshift range of $ 0.001 < z < 0.08$ with a median redshift of 0.02 and a total stellar mass and SFR range of $7.7 < \log(M_{\star}/\textrm{M}_{\odot}) < 11.7$ and $ -2.2 < \log(\textrm{SFR}/\textrm{M}_{\odot}\textrm{yr}^{-1}) < 2.1 $ respectively. These mass and SFR ranges are at least 1 dex broader than previous studies of the SGMS in local galaxies.	

\subsection{IRAC Matched Aperture Photometry} \label{Analysis}

We based our analysis of the total and spatially resolved MS (Sections \ref{globalMS} and \ref{SGMS}) on SFRs and stellar masses estimated with the IRAC 8.0 and 3.6 $\micron$ flux densities tabulated by \cite{SFRS}. The 8.0 $ \micron $ band provides better resolution for measuring SFR compared to MIPS 24 $ \micron $, and it is relatively unaffected by absorption. Numerous studies have shown that in metal-rich SFGs, polycyclic aromatic hydrocarbon (PAH) luminosity scales relatively well with SFR (\citealt{CalzettiPAH11}; \citealt{KE12} (KE12) and references within). Because the majority of SFRS SFGs have metallicities close to solar (Maragkoudakis et al. in preparation), the 8.0 $\micron$ luminosity should be a good SFR tracer.

The initial reduction of the IRAC data and construction of image mosaics was described by \cite{SFRS}. To measure the 3.6 and 8.0 $ \micron $ flux on sub-galactic scales, we constructed a grid of square apertures centered on the nucleus, based on the source coordinates (\citealt{SFRS}). Each aperture corresponds to a physical size of $1 \times 1$ kpc$^{2}$. The angular aperture sizes in each galaxy were adjusted according to galaxy distance to ensure consistent sampling of linear scales throughout our study sample. This choice of angular scale was driven by three considerations: (1) 1\,kpc is the minimum length scale on which the standard SFR metrics are thought to be reliable (e.g., KE12); (2) we require multiple regions within each galaxy; (3) the IRAC 8.0 $ \micron $ angular resolution of 2\arcsec corresponds to 1kpc at a distance of 100 Mpc and 70\% of the SFRS SFGs are closer than this.

\begin{figure}
	\begin{center}
		\includegraphics[keepaspectratio=true, scale=.36]{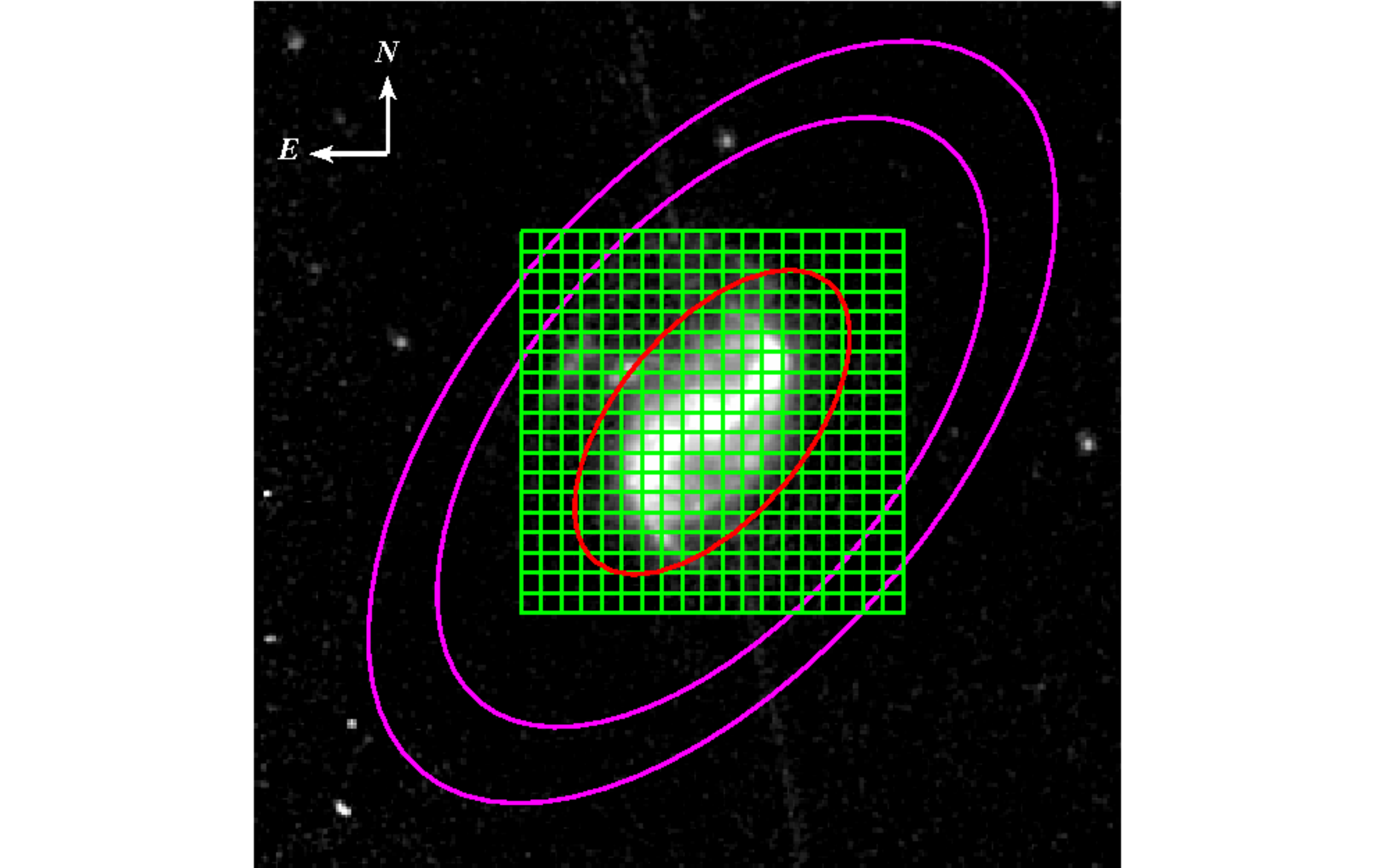}
		\caption{Example of IRAC 8.0 $\micron$ multiple aperture photometry on galaxy NGC~2718. Small green boxes correspond to a physical area of $1 \times 1$ kpc$^{2}$. Red ellipse corresponds to 1.5 R$_{{\textrm{P}90}}$ and magenta ellipses to 2R$_{{\textrm{P}90}}$ and 2.5R$_{{\textrm{P}90}}$. The photometry was performed in boxes within the smallest red ellipse. Sky subtraction was performed based on the median background measured between the outer ellipses.}
		\label{Grid}
	\end{center}
\end{figure}

We mapped sub-galactic regions within 1.5 times each galaxy's Petrosian radius R$_{{\textrm{P}90}}$, defined as the radius including 90\% of the galaxy light (Figure \ref{Grid}). Because we are interested in star-forming regions, we rejected regions having surface brightness smaller than 10\% of the second highest surface brightness region at 8.0 $ \micron $. The median galaxy had 16 apertures with minimum 1 aperture (11 galaxies) and maximum 452 apertures (1 galaxy). The photometry was performed with CIAO's (\citealt{CIAO}) ``dmstat" and ``dmextract" tools. The background contribution was subtracted based on the median value of the background within an elliptical annulus centered on the galaxy and having inner and outer radii of 2 and 2.5 times the galaxy's R$_{\textrm{p}_{90}}$ respectively. The orientation and ellipticity of the background region, in each case, matched that of the galaxy.

\subsection{Mass and Star-Formation Rate Estimations} \label{Mass-SFR}

Total stellar masses and stellar mass in sub-galactic apertures were calculated using the \cite{Zhu10} mass-to-light ratio calibration for the IRAC 3.6 $\micron$ band: \begin{equation} \label{eq:Stellar-Mass} \frac{M_{\star}}{M_{\odot}} = 10^{0.23+(1.14)(g-r)} \times \frac{\nu L_{\nu3.6\,\micron}}{L_{\odot}}\end{equation} where $g$, $r$ are the Petrosian AB magnitudes obtained from the SDSS DR12 (\citealt{SDSSDR12}), and galaxy 3.6 $ \micron $ magnitudes are from \cite{SFRS}. Subregion $ g-r $ colors were assumed to be the same as the whole galaxy. For the nuclear regions, stellar populations in general consist of a mixture of old stars from the bulge and young stars in circumnuclear star-forming regions. We used the SDSS $g-r$ fiber colors for the nucleus. 

\begin{figure*}
	\begin{center}
		\includegraphics[keepaspectratio=true, scale=.55]{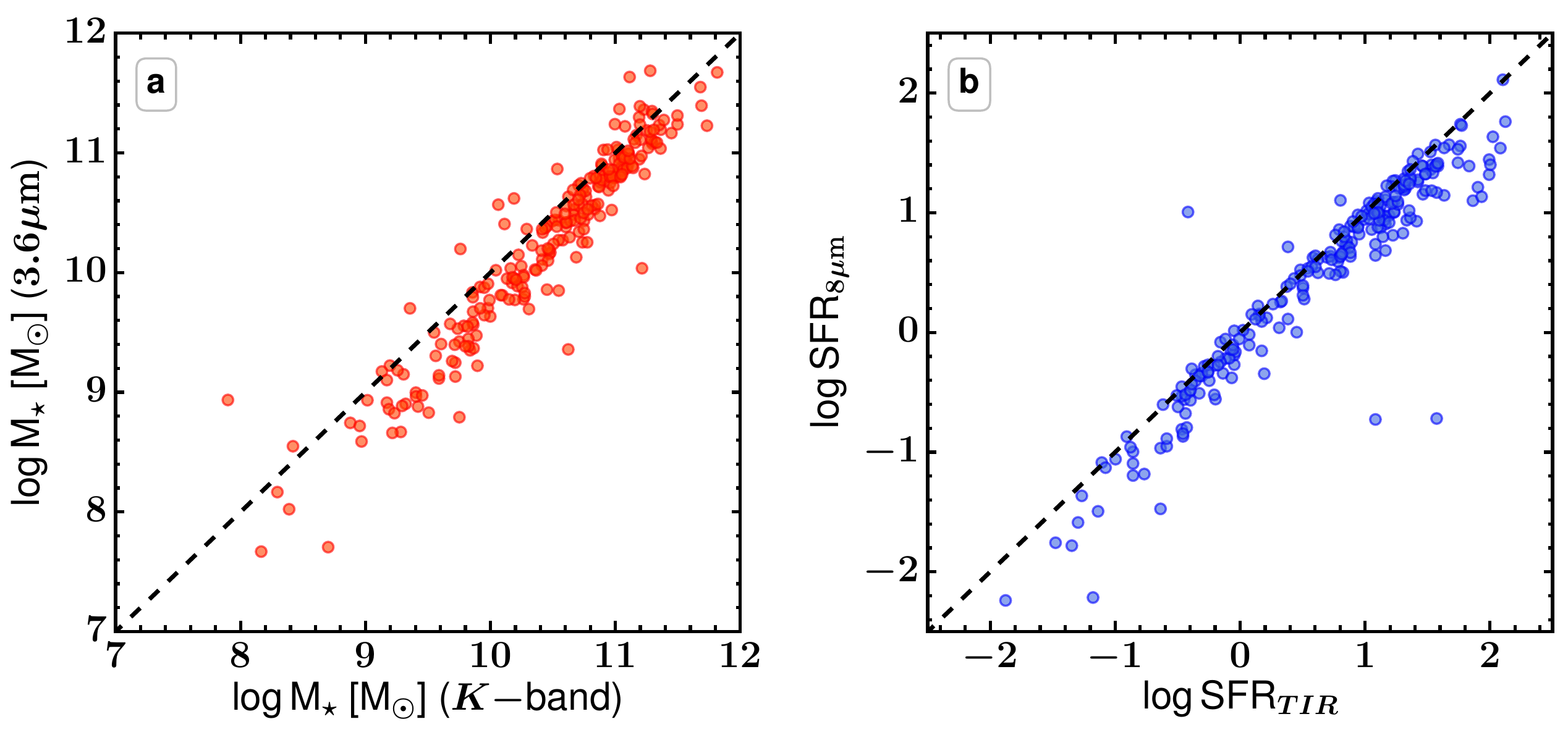}
		\caption{A comparison of the total galaxy SFR and stellar mass estimators. Left panel: Total galaxy stellar mass estimated from the 3.6 $\micron$ calibration (Eq. \ref{eq:Stellar-Mass}) plotted against the $K$-band calibration (Eq. \ref{eq:MassKs}). Right panel: Total SFR estimated from the 8.0 $\micron$ calibration (Eq. \ref{eq:SFRI4}) plotted against the TIR calibration (Eq. \ref{eq:SFRTIR}). Dashed equality lines are drawn in both panels.}
		\label{tracers}
	\end{center}
\end{figure*}

We verified the 3.6 $\micron$ derived stellar masses by comparing them with the commonly used and well-calibrated $K$-band relation described by \cite{Bell03}: \begin{equation} \label{eq:MassKs} \frac{M_{\star}}{M_{\odot}} = 10^{-0.273+(0.091)(u-r)} \times \frac{L_{K_{s}}}{L_{K_{\odot}}}\end{equation} where $u-r$ are the Petrosian SDSS colors in AB magnitudes. We calculated $K$-band luminosities using the asymptotic $K$-band magnitudes measured from integrating the 2D model fits to the 2MASS images of the galaxies (Bonfini et al. in preparation). Figure \ref{tracers}a shows the comparison. The 3.6 $\micron$ tracer shows a systematic underestimation of the stellar-mass compared to the $K$-band calibration. The integrated 3.6 $ \micron $ photometry was measured by using SExtractor (\citealt{SFRS}), which is subject to aperture effects depending on the galaxy light traced. The asymptotic $K$-band fluxes measured from fitting the galaxies profiles by Bonfini et al. are generally larger than the ones derived by SExtractor (\citealt{SFRS}). Overall the two stellar-mass tracers differ simply by an offset without showing non-linearity effects, and the offset does not influence the analysis.

We calculated total SFRs and SFRs in sub-galactic apertures from the starlight-subtracted 8.0 $\micron$ images. The starlight subtraction was performed using the scaling of \cite{Helou04}: \begin{equation} f_{8\,\micron ,(\textrm{PAH})} = f_{8\,\micron} - 0.26 f_{3.6\,\micron} \end{equation} Then the SFR was calculated using the calibration of \cite{Pancoast10}\footnote{In principle the SFR should include ultraviolet emission, but for the SFRS sample, this is typically a 15\% effect and never as much as a factor of 2 (Mahajan et al., in preparation).}: \begin{equation} \label{eq:SFRI4} \frac{\textrm{SFR}_{8\,\micron}}{[\textrm{M}_{\odot}/\textrm{yr)}]} = 6.3 \times 10^{-10} \frac{L_{8\,\micron}}{L_{\odot}} \end{equation} For comparison purposes we also measured the total SFR from the TIR luminosity (\citealt{SFRS}) using the \cite{DH02} TIR flux formula $F_{TIR} = 2.403\nu f_{\nu25\,\micron} - 0.245\nu f_{\nu60\,\micron} + 1.347\nu f_{\nu100\,\micron}$ but using the MIPS 24\,\micron\ flux densities where available instead of the less precise IRAS 25\,\micron\ flux densities. We then calculated the SFR using the TIR calibration by KE12: \begin{equation} \label{eq:SFRTIR} \frac{\textrm{SFR}}{[\textrm{M}_{\odot}/\textrm{yr)}]} = \frac{L_{\textrm{TIR}}}{[\textrm{erg/s}]} \times 10^{-43.41}
\end{equation} Figure \ref{tracers}b compares the 8.0~$ \micron $ and TIR SFR tracers. The two SFR tracers agree apart from an offset, with the 8.0~$ \micron $ underestimating the SFR compared to the TIR which is expected because the TIR also traces the contribution from low-mass stars, as well as encompasses the systematic error from the contribution of evolved populations (above $ \sim $200 Myr) leading to an overestimate of SFR. As with the different stellar mass estimators, there is only an offset between the two tracers in an otherwise linear conversion.

\section{The Global Main Sequence of Local Star-Forming Galaxies} \label{globalMS}
\begin{figure}
	\begin{center}
		\includegraphics[keepaspectratio=true, scale=.55]{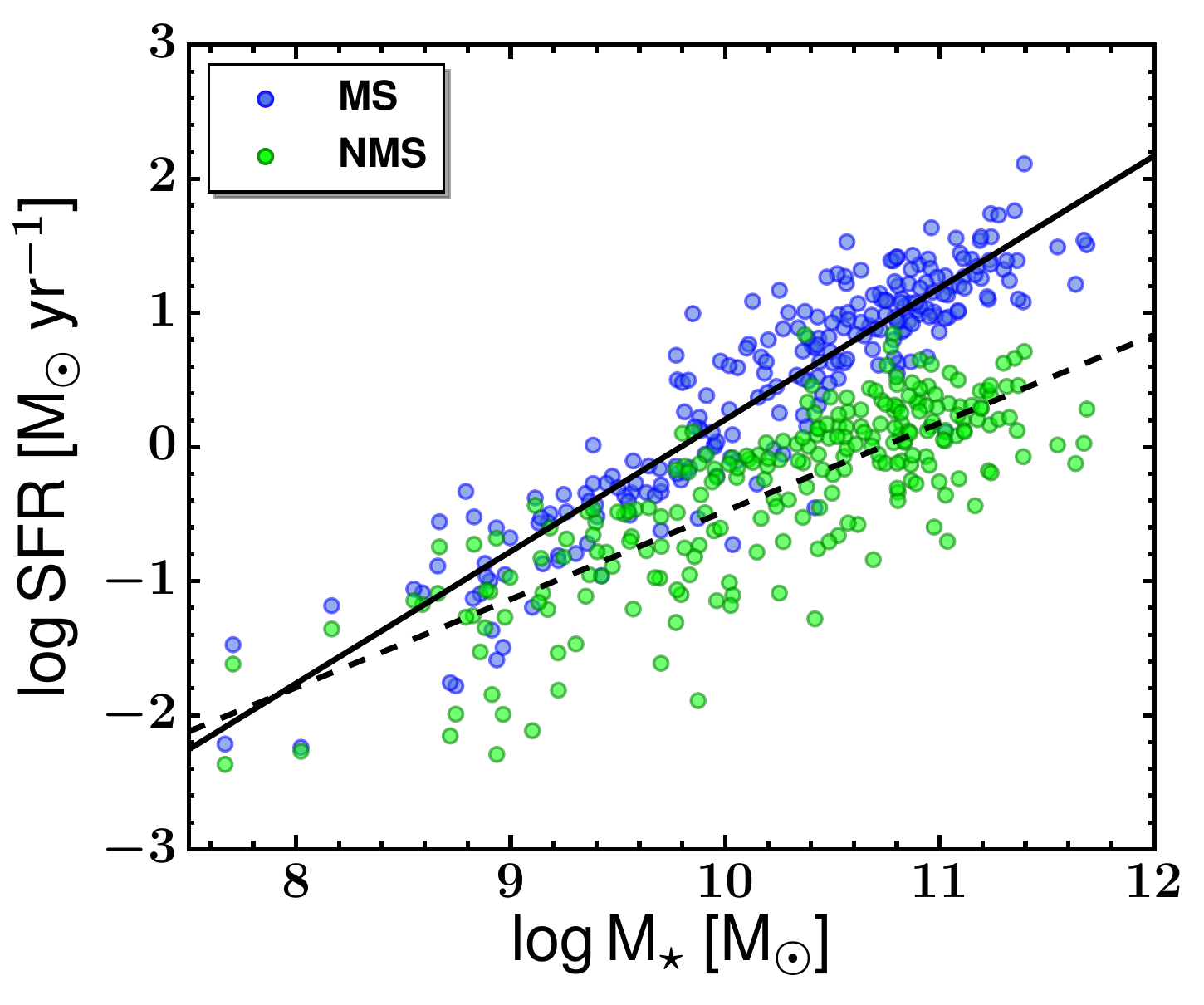}
		\caption{The global MS (blue points) and nuclear region MS (NMS) (green points) of the SFRS SFGs. The SFR is calculated using the 8 $\micron$ calibration, and stellar mass is derived from the 3.6 $\micron$ tracer. Both SFRs are plotted against total stellar mass. The solid line is the fit to the MS (Eq. \ref{eq:MSfit}), and the dashed line is the fit to the NMS. The properties of the two sequences are given in Table \ref{ms-info}.}
		\label{MS-NMS}
	\end{center}
\end{figure}

With a wide range of galaxy parameters and a fully representative star-forming galaxy sample, the SFRS SFGs provide an ideal set to examine the MS plane in the local Universe, both in integrated and sub-galactic scales. Figure \ref{MS-NMS} shows the standard MS fitted with a linear function, in log space, of the form: \begin{equation}  \label{eq:MSfit}
\log \textrm{(SFR/M}_{\odot}\textrm{yr}^{-1}) = \alpha \log (M_{\star}/\textrm{M}_{\odot}) + \beta
\end{equation} The best-fit parameters (Table \ref{ms-info}) show that the MS of our sample of local galaxies has a slope of $ \alpha = 0.98$ and a normalization of $-9.63 \log(\textrm{M}_{\odot}\textrm{yr}^{-1})$. The dispersion about the best-fit line is 0.32\,dex, consistent with the $\sim$0.35~dex MS scatter reported in the literature (S14 and references within).

\begin{table*}
	\begin{minipage}{125mm}

		\caption{MS, NMS, and SGMS best-fit parameters.}
		\label{ms-info}
		\begin{tabular}{@{}cccccc}
			\hline
			Sequence & Num. of galaxies & Apertures & $\alpha$ & $\beta$ & $\sigma$ \\
			\hline
			MS & $245^{a}$ & - & 0.98 $\pm$ 0.03 & $-$9.63 $\pm$ 0.27 & 0.32 \\
			NMS & 246 & - & 0.66 $\pm$ 0.03 & $-$7.05 $\pm$ 0.33 & 0.39 \\
			SGMS & 246 & 16 & 0.91 $\pm$ 0.01 & $-$9.01 $\pm$ 0.05 & 0.31 \\
			All morphological types SGMS & 130 & 14 & 0.89 $\pm$ 0.01 &  $ -8.97 \pm  0.07$ & 0.31 \\
			(E--S0/a) SGMS & 10 & 11 & 1.09 $\pm$ 0.03 & $-$10.76 $\pm$ 0.22 & 0.18 \\
			(Sa--Sbc) SGMS & 55 & 17 & 0.81 $\pm$ 0.02 & $-$8.30 $\pm$ 0.13 & 0.30 \\
			(Sc--Irr) SGMS & 65 & 15 & 0.97 $\pm$ 0.01 & $-$9.52 $\pm$ 0.09 & 0.30 \\		
			Unclassified galaxies SGMS & 116 & 19 & 0.72 $\pm$ 0.01 & $-$7.25 $\pm$ 0.08 & 0.26 \\
			\hline
		\end{tabular}

	\medskip	
	Fits of the form $\log SFR = \alpha \log M_{\star} + \beta$ to the MS in different scales. Column (1): The various sequences examined; Column (2): the number of galaxies in each sequence; Column (3): the median number of apertures per galaxy; Column (4) the best-fit slope ($\alpha$); Column (5): the zero-point ($\beta$) value of the best fit; Column (6) the metric of dispersion ($\sigma$) in dex from the best-fit line.\\
	$ ^{a} $ Galaxy IC~3476 has no available total 8.0 $\micron$ photometry.
\end{minipage}
\end{table*}

\section{The sub-galactic Main Sequence of Star-Forming Galaxies} \label{SGMS}

Three variants of the SGMS comprise sequences based on (1) all sub-galactic regions identified in Section \ref{Analysis}; (2) bright regions and random regions, excluding the nuclei; (3) only the nuclear regions.

\subsection{The $\Sigma_{\textrm{SFR}}$ -- $\Sigma_{\star}$ Correlation} \label{SGMS-all}

\begin{figure*}
	\begin{center}
		\includegraphics[keepaspectratio=true, scale=.54]{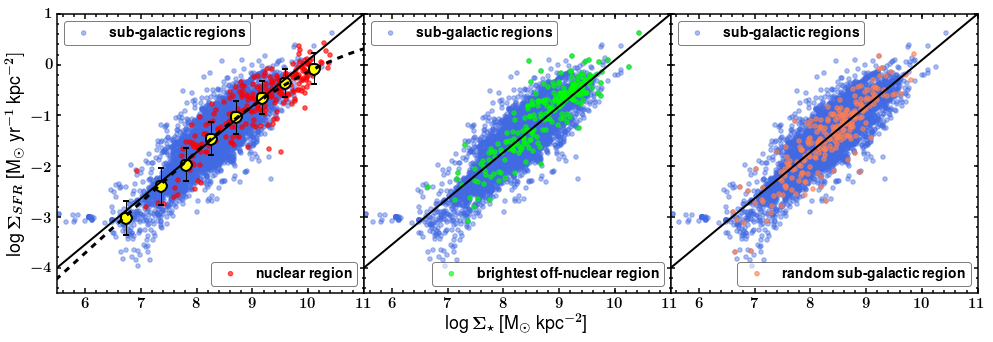}
		\caption{The sub-galactic main sequence between $\Sigma_{\textrm{SFR}}$ and $\Sigma_{\star}$ for the 246 star-forming galaxies in the SFRS. The blue points in all panels represent the individual sub-galactic regions of all SFGs. The red points in the left panel present the sequence of the nuclear regions, and the yellow points show the mean SFR density in mass bins. The first mass bin is defined between 5.5--7.0 $\log\Sigma_{\star} [\textrm{M}_\odot $kpc$ ^{-2}]$ and all later at every 0.5 dex mass. The green points in the middle panel show the off-nuclear regions of all galaxies with the highest surface brightness, and the orange points in the right panel correspond to randomly drawn sub-galactic regions. The best-fit line of the SGMS is shown with a solid line in all panels. A $ M_{\star}$-dependent fit for all the regions based on the parametrization of \protect\cite{Schreiber15} is shown with a dashed line in the left panel.}
		\label{SGMS3panel}
	\end{center}
\end{figure*}

In all variants of the SGMS we measured the star formation surface density $\Sigma_{\textrm{SFR}}$ defined as:
\begin{equation}
\Sigma_{\textrm{SFR}} = \textrm{SFR/Area} \quad [\textrm{M}_{\odot} \textrm{yr}^{-1} \textrm{kpc}^{-2}]
\end{equation}
and similarly the stellar mass surface density ($\Sigma_{\star}$). Figure \ref{SGMS3panel} shows the result, revealing a strong correlation between these two parameters for regions on physical scales of 1 kpc. In all cases we fitted the relation with a linear function in logarithmic space. We performed an additional $ M_{\star} $-dependent fit using the \cite{Schreiber15} (S15) eq. 9 parametrization, in order to examine whether the flattening towards higher stellar densities reported by \cite{Wuyts13} is present in our sample. We set all $z$-dependent terms to zero, because all SFRS galaxies allocate a narrow redshift range, but allowed the remaining three parameters to vary. The fit (Figure \ref{SGMS3panel}) is slightly better in the highest mass bin, but the decrease in overall scatter is only from 0.310~dex to 0.308~dex. The rest of this paper discusses only the linear fit for simplicity and easier comparison to previous results. The mean SFR density in mass bins (Figure \ref{SGMS3panel}) is consistent with the best-fit line within the measured uncertainties. The scatter in the SGMS is 0.31 dex, similar to the 0.32 dex scatter of the global MS and within the range of scatter reported in different MS studies (S14). The best-fit values are shown in Table \ref{ms-info} along with the dispersion ($\sigma$). The SGMS slope is $ \alpha_{\textrm{SGMS}} = 0.91$, slightly less than that of the integrated MS ($ \alpha_{\textrm{MS}} $ = 0.98). 

\begin{figure*}
	\begin{center}
		\includegraphics[keepaspectratio=true, scale=.45]{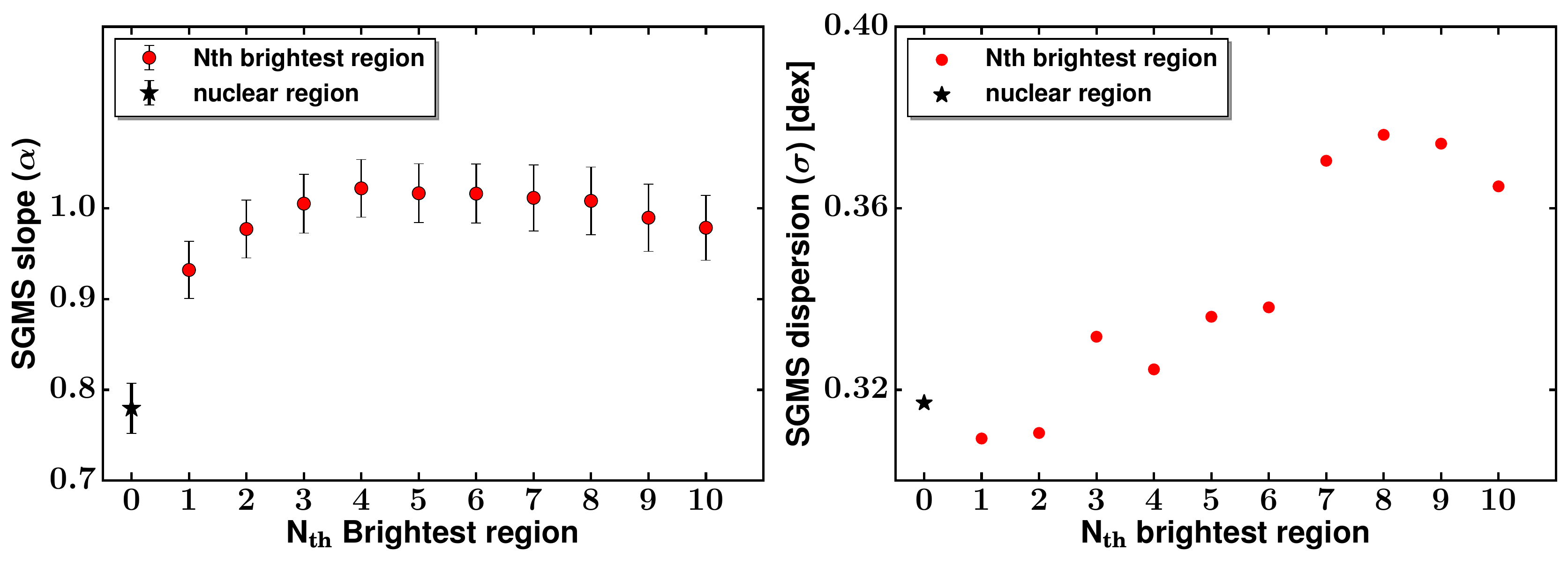}
		\caption{The slope (\textbf{left panel}) and dispersion (\textbf{right panel}) of the SGMS produced individually from reduced regions of the highest $ 8\micron $ surface brightness. Index 0 corresponds to galaxy nuclei, index 1 to the brightest off-nuclear regions, and 10 to the 10th brightest off-nuclear region of each galaxy. The dispersion and slope for the SGMS of the nuclear regions are shown with a star symbol (region number 0).}
		\label{Nth-region}
	\end{center}	
\end{figure*}

In order to determine whether the SGMS, and possibly the global MS, are driven by the most active star-forming regions in the galaxy, we measured the SGMS parameters formed from a single region in each galaxy, beginning from the brightest off-nuclear region and progressing to regions of lower surface brightness in the 8 $\micron$ IRAC band. For this analysis we excluded the nuclei (which are discussed in Section \ref{nucleus}) and concentrated explicitly on star formation taking place in the disc. For the 236 galaxies covered by one or more apertures outside the nucleus, we measured $\Sigma_{\textrm{SFR}}$ and $\Sigma_{\star}$ of the brightest off-nuclear region and calculated the slope and dispersion of the produced SGMS. Then, we repeated this analysis for the next brightest region for each galaxy down to the 10th in line. Figure \ref{Nth-region} presents the results. The slope of the SGMS of different (nth) surface brightness regions asymptotically increases towards the slope of the overall SGMS, with regions of higher surface brightness having shallower slopes. Therefore, higher surface brightness regions are not the primary drivers of the correlation. In addition, the right panel of Figure \ref{Nth-region} shows that the measured SGMS scatter for regions of different surface brightness increases progressively from regions of higher to lower surface brightness. This is indicative that the MS and SGMS are not governed by the most active sub-galactic regions in the galaxy.

Furthermore, in order to investigate whether sampling one individual region from different galaxies will give a good representation of the $\Sigma_{\textrm{SFR}}$--$\Sigma_{\star}$ correlation, we measured the SGMS slope and dispersion produced from random sub-galactic regions. Specifically, we randomly drew a single region from each galaxy and measured the dispersion and slope of the produced SGMS, repeating the process 1000 times. The random-region SGMS from one such draw is shown in Figure \ref{SGMS3panel}. The dispersion in the 1000 trials varied between 0.31 and 0.38 dex with a median of 0.34 dex, and the slope ranged between values of 0.86 and 1.03 with a median of 0.95. This is consistent with our previous result, considering all regions as well as regions of different surface brightness in each galaxy, showing that scatter is driven by the general sub-galactic populations rather than the less active. It also suggests that most sub-galactic star-forming regions can be representative of the star-forming conditions overall.

\begin{figure}
	\begin{center}
		\includegraphics[keepaspectratio=true, scale=.54]{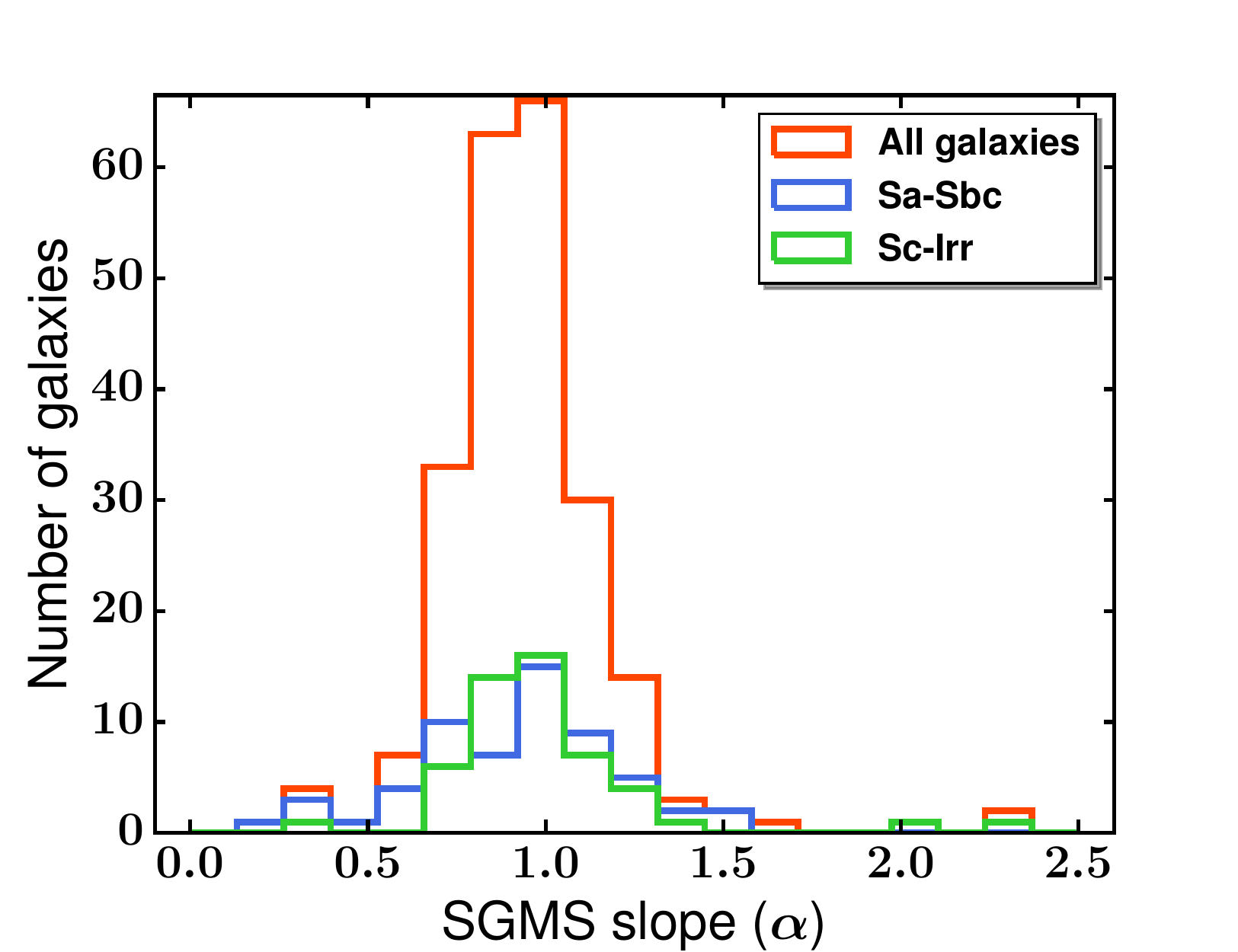}
		\caption{The distribution of SGMS slopes produced by distinct regions within individual galaxies. The histogram for all galaxies is shown in red color, the ``early-type spirals" group histogram is shown in blue, and the ``late-type" group in green.}
		\label{islopes}
	\end{center}
\end{figure}

\begin{figure*}
 	\begin{center}
 		\includegraphics[keepaspectratio=true, scale=.54]{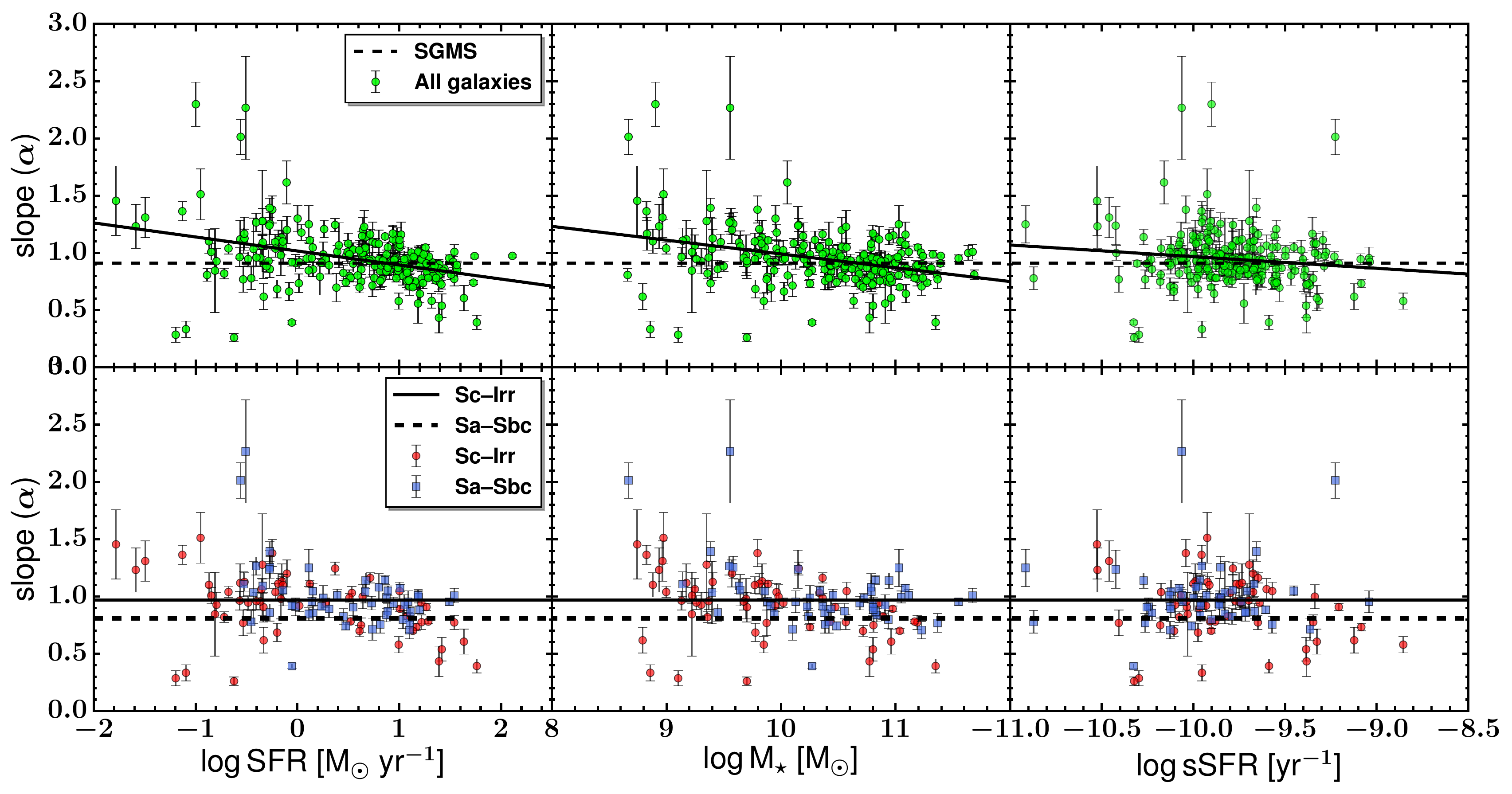}
 		\caption{The SGMS slopes within individual galaxies plotted against total SFR (left panels), total M$ _{\star} $ (middle panels), and sSFR (right panels). The top three panels show all the galaxies used (with or without morphological classification), while the bottom three panels show galaxies in the ``late-types" group (red circles) and the ``early-type spirals" group (blue boxes). In the top panels the SGMS slope of all galaxies is shown with a dashed line, while the best-fit line in each case is shown with a solid line. In the bottom panels the SGMS  slope of the ``early-type spirals" group is shown with a dashed line and the SGMS slope of the ``late-types" group with a solid line.}
 		\label{islope_2x3panel}
 	\end{center}
 \end{figure*}

To trace the behavior of the SGMS in individual galaxies and examine how well it resembles the global SGMS produced from all sub-regions of the entire SFG sample, we measured the SGMS slope, dispersion, and the local sSFR from individual galaxies. In order to obtain meaningful constraints on the SGMS parameters, we examined the 230 galaxies that are resolved in at least 5 apertures. Figure \ref{islopes} shows the distribution of individual SGMS slopes.The distribution peaks at a mean value of 0.95 very close to the global SGMS slope ($ \alpha_{\textrm{SGMS}} = 0.91$). Within a single galaxy, stars form largely where the stellar mass is already high. Figure \ref{islope_2x3panel} shows the variation of the individual galaxy SGMS slopes with respect to their total SFR, total M$ _{\star} $, and sSFR. The fits reveal a tendency of the individual SGMS slopes to decrease towards higher values of total SFR and total M$ _{\star} $. Most galaxies at $\log(\textrm{SFR}/\textrm{M}_{\odot} \textrm{yr}^{-1}) < 0 $ and $\log(\textrm{M}_{\star}/\textrm{M}_{\odot}) < 10$ show a super-linear ($ \alpha > 1 $) correlation between SFR density and stellar-mass density, and the opposite behavior, i.e. a sub-linear correlation, for larger total SFR and M$ _{\star} $ values. Nevertheless, the correlation between local mass and local SFR holds for galaxies with a wide range of properties.

\begin{figure}
	\begin{center}
		\includegraphics[keepaspectratio=true, scale=.5	]{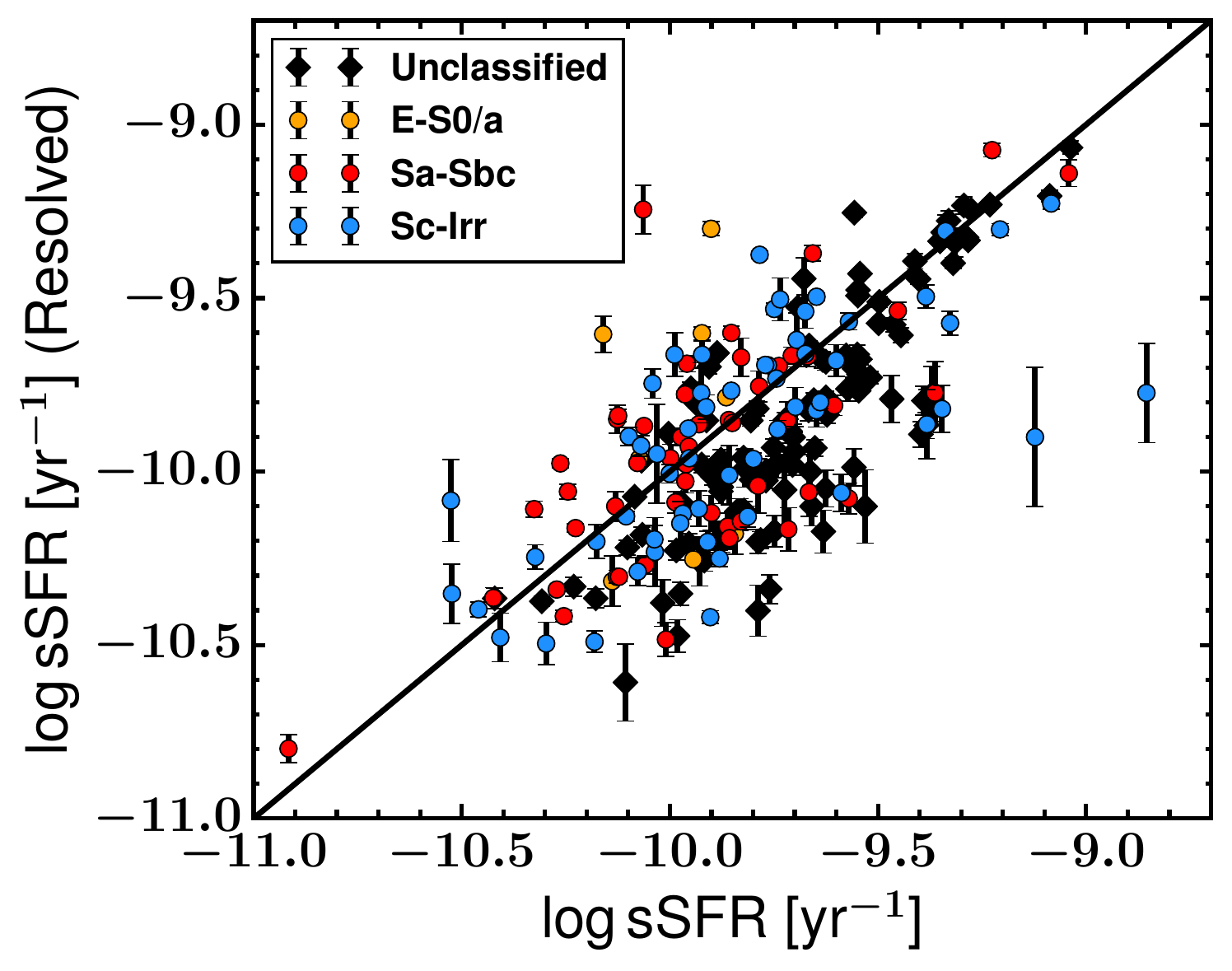}
		\caption{Comparison of sSFRs for individual galaxies. The abscissa shows the global sSFR. The ordinate shows the sSFR measured from the slope derived from a linear fit to each galaxy's SGMS. The solid line indicates equality. Points are color-coded by galaxy type as indicated.}
		\label{isSFR_vs_sSFR}
	\end{center}
\end{figure}

Figure \ref{isSFR_vs_sSFR} shows how a local average of the sSFR measured from the sub-galactic regions of a galaxy correlates to the ratio of its total sSFR. The total sSFR is the total SFR divided by total stellar mass. The local average sSFR, plotted on the vertical axis of Figure \ref{isSFR_vs_sSFR}, was calculated from the fitted slope in the non-logarithmic correlation $\Sigma_{\rm SFR} = \alpha \Sigma_* + \beta$ of the sub-galactic regions in each galaxy. The two measurements agree within 0.3 dex scatter. In other words, galaxies with high sSFR show elevated sSFR on average everywhere (at $ \sim $1 kpc spatial scale), not just in selected regions, and SFR correlates with mass at kpc scales in individual galaxies.
 
\subsection{Morphology Dependence of the SGMS} \label{morphology}

While the morphological dependence of the galaxy MS has been examined (\citealt{Noeske07}, \citealt{Wuyts11}), relatively little is known about the morphological dependence of the SGMS. We investigated the SGMS for a subset of 130 SFRS SFGs with available morphological classification (\citealt{SFRS}). We defined three morphological groups. The ``lenticulars" group (10 galaxies) contains the single elliptical galaxy in our sample along with S0 to S0/a types. The ``early-type spirals" group (55 galaxies) consists of Sa to Sbc types, and lastly all types Sc and later (65 galaxies) are in the ``late-type" group. A fourth group containing the galaxies with unknown morphological classes is also defined and studied separately. 

\begin{figure*}
	\begin{center}
		\includegraphics[keepaspectratio=true, scale=.58]{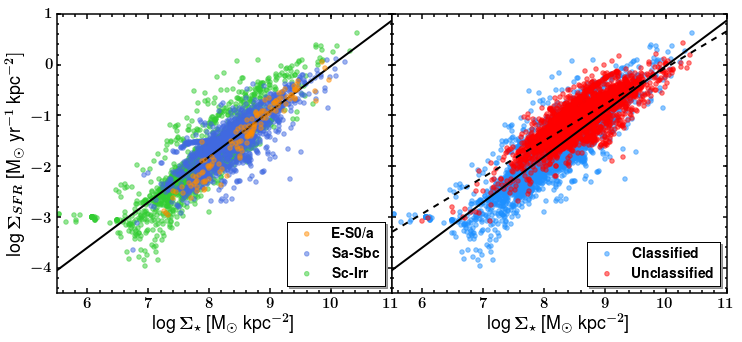}
		\caption{\textbf{Left panel}: The sub-galactic main sequence produced from all morphologically classified galaxies. E to S0/a Hubble types are shown with orange points, Sa--Sbc with blue points, and all types Sc and later with green. \textbf{Right panel}: The sub-galactic main sequence of all SFGs with unknown morphological classifications (red points) overplotted on the SGMS of galaxies with known morphological types (blue points). The dotted line is the best fit for galaxies with unknown morphologies, and the solid line in both panels is the best-fit for the combined morphological groups.}
		\label{SGMS_All_Morph}
	\end{center}
\end{figure*}

\begin{figure*}
	\begin{center}
		\includegraphics[keepaspectratio=true, scale=.5]{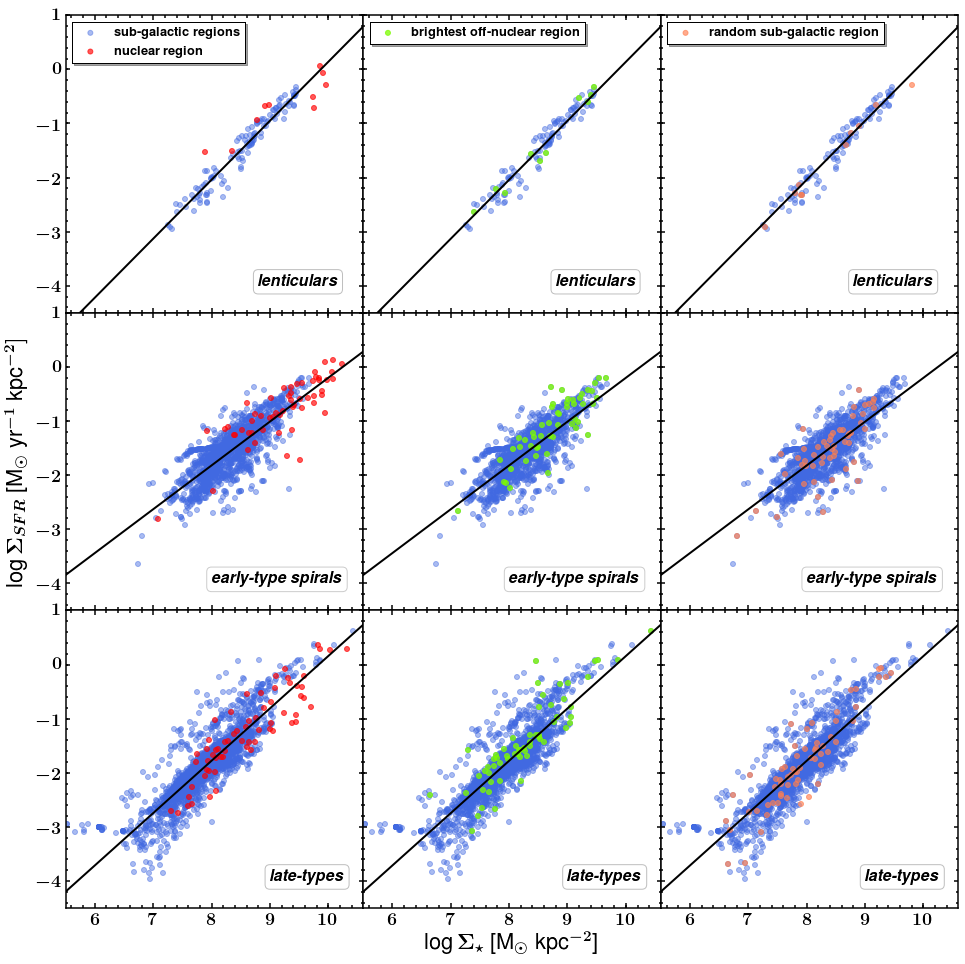}
		\caption{The sub-galactic main sequences calculated for different morphological groups (Top row: E to SO/a, middle row: Sa to Sbc, bottom row: Sc and later). The blue points in all panels represent all the sub-galactic regions from the galaxies in the group shown. The nuclear regions are indicated with red points in the left panels, the brightest off-nuclear regions from each galaxy are shown with green points in the middle panels, and randomly-drawn regions from each galaxy are shown with orange points in the right panels. The best fit to the SGMS (blue points) is shown with black lines.}
		\label{SGMS-morph}
	\end{center}
\end{figure*}

The SGMS of all classified morphological types gives the exact same slope and dispersion as the ``all-galaxies" SGMS (Figure \ref{SGMS_All_Morph}). The corresponding SGMS of morphologically unclassified galaxies has a sub-linear slope. Figure \ref{SGMS-morph} shows the sub-galactic $\Sigma_{\textrm{SFR}}$ against $\Sigma_{\star}$ for all regions of the three morphological groups, and the fit parameters are given in Table \ref{ms-info}. The ``lenticular" and ``late-type" groups have the steepest correlation between $\Sigma_{\textrm{SFR}}$ and $\Sigma_{\star}$. In addition, Figures \ref{islopes} and \ref{islope_2x3panel} show the distributions of the individual galaxy SGMS slopes and their variation with total SFR, total M$ _{\star} $, and sSFR separated by morphology. Overall there is no substantial variation between the previous properties and the individual SGMS slopes with morphology.

\subsection{The Nuclear Region And Nuclear Main Sequence} \label{nucleus}

\begin{figure*}
	\begin{center}
		\includegraphics[keepaspectratio=true, scale=.45]{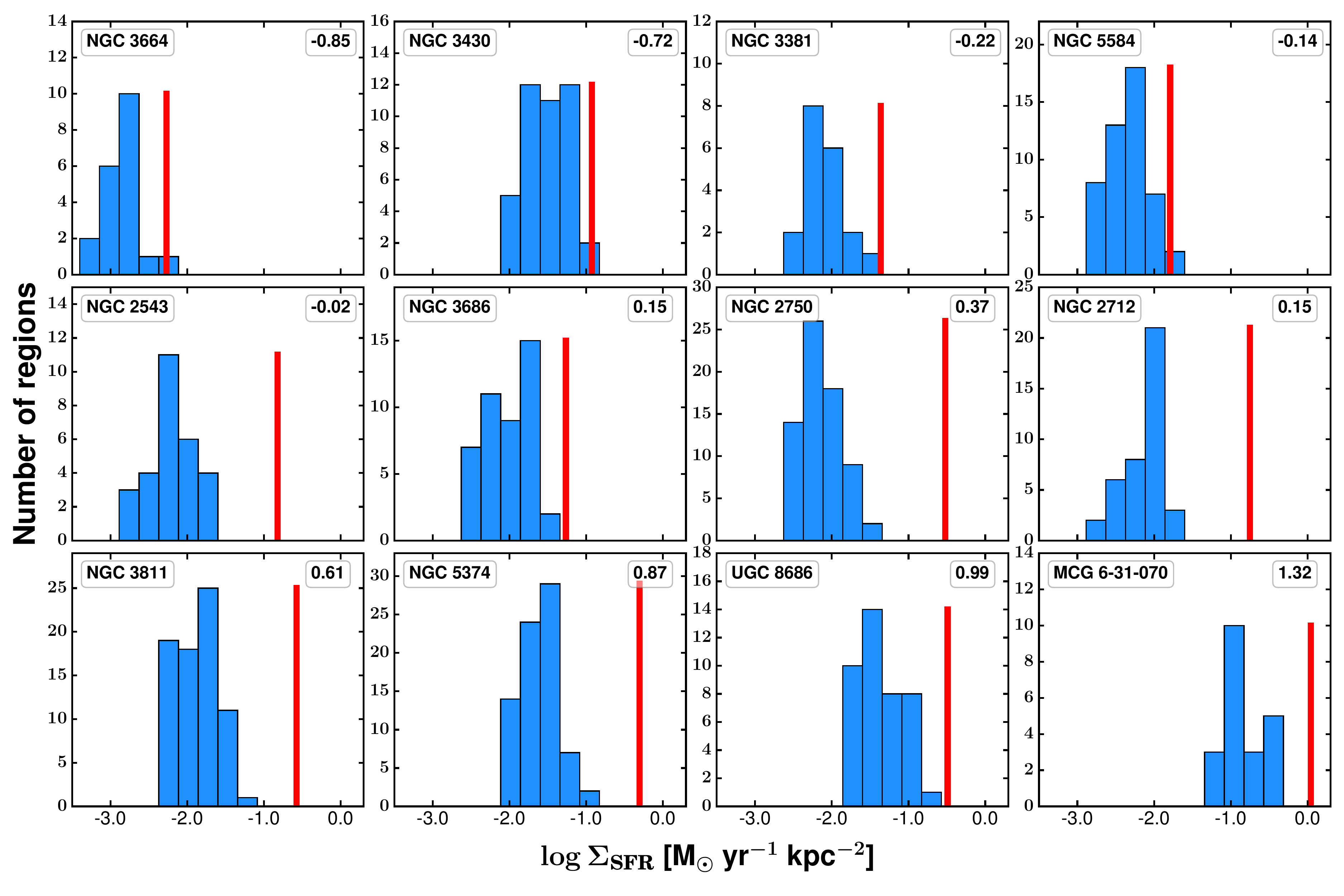}
		\caption{The histograms of off-nucleus $\Sigma_{\textrm{SFR}}$ within 12 galaxies of different SFR (0.14 -- 20.9 M$_{\odot}$/yr) and stellar mass content ($10^{9.05} - 10^{11.17}$ M$_{\odot}$), chosen to have a large number of apertures ($>15$). The red bars indicate $\Sigma_{\textrm{SFR}}$ of the nuclear region. The total $ \log\textrm{SFR} $ of each galaxy is given at the top right corner of each panel in units of $\log$ [M$_{\odot}$ yr$^{-1}$].}
		\label{SFR_hist}
	\end{center}
\end{figure*}

\begin{figure*}
	\begin{center}
		\includegraphics[keepaspectratio=true, scale=.45]{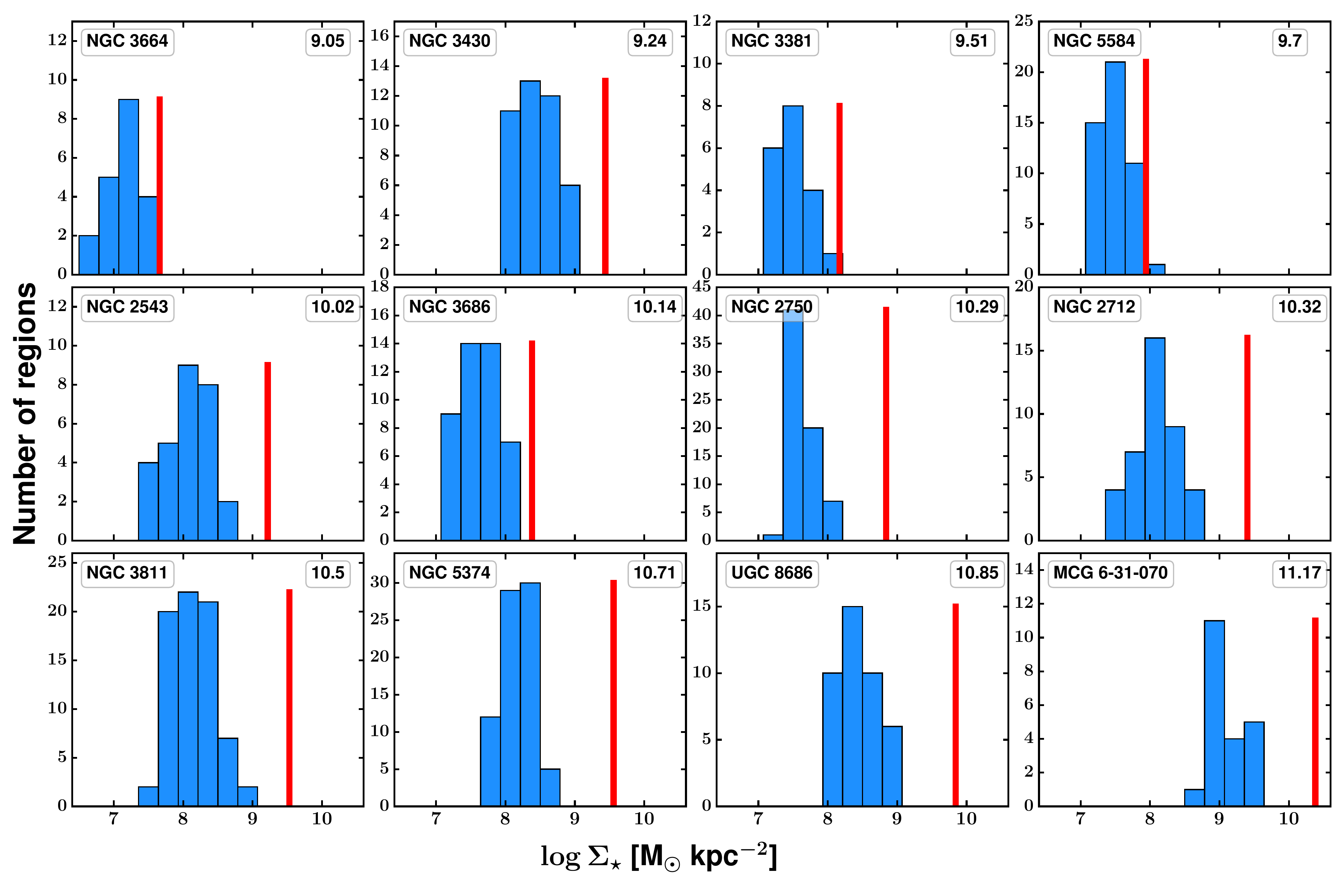}
		\caption{The histograms of $\Sigma_{\star}$ for off-nucleus regions of the 12 galaxies presented in Figure \ref{SFR_hist}. The red bars indicate $\Sigma_{\star}$ of the nuclear region. The total stellar mass of each galaxy is given at the top right corner of each panel in units of $\log(\textrm{M}_{\star}/\textrm{M}_{\odot})$.}
		\label{Mass_hist}
	\end{center}
\end{figure*}

Because galaxy nuclei may contain supermassive black holes, these regions can host processes other than star formation (e.g. accretion, outflows, intense UV radiation). The (circum-) nuclear regions hold the most intense activity of star formation in galaxies as well as being the most massive in stellar content for nearly all galaxies in the sample, regardless of Hubble type (Figure \ref{SGMS-morph}). For the 236 galaxies with more than the nuclear aperture covering their surface, the nucleus is the maximum SFR and mass densities for at least 93\%\footnote{There are 27 galaxies where the adopted nucleus is not the brightest spot. Of these, 11 have wrong nominal coordinates for the nucleus, which is in fact the brightest spot in the galaxy. The remaining 16 have ambiguous or undefined nuclei because of low surface brightness, multiple nuclei, or possible field stars.}. The examples in Figures \ref{SFR_hist} and \ref{Mass_hist} are typically showing that the nuclear $\Sigma_{\textrm{SFR}}$ and $\Sigma_{\star}$ are among the highest values in the galaxies and in most cases are the maximum values. Because the average effective bulge radii for all SFRS galaxies is $\sim$1.4~kpc (Bonfini et al. in preparation) and considering that the aperture size in our analysis is 1 kpc, our nuclear apertures likely include emission from circumnuclear star-forming activity. For the same reason, apertures adjacent to the nucleus may contain some bulge light which would reduce the inferred sSFR in these regions. Furthermore, in cases of irregular galaxies with no definable nucleus or multiple galaxy systems (e.g., mergers), the coordinates may not target precisely the nucleus. However, 75\% of the brightest off-nuclear regions are 1 kpc from the nucleus, 18\% lie at distances between 2 and 4 kpc, and only 7\% are $ >4 $ kpc. This confirms the general trend that the circumnuclear area is where star-formation activity and stellar mass concentration peak.

\begin{figure}
	\begin{center}
		\includegraphics[keepaspectratio=true, scale=.52]{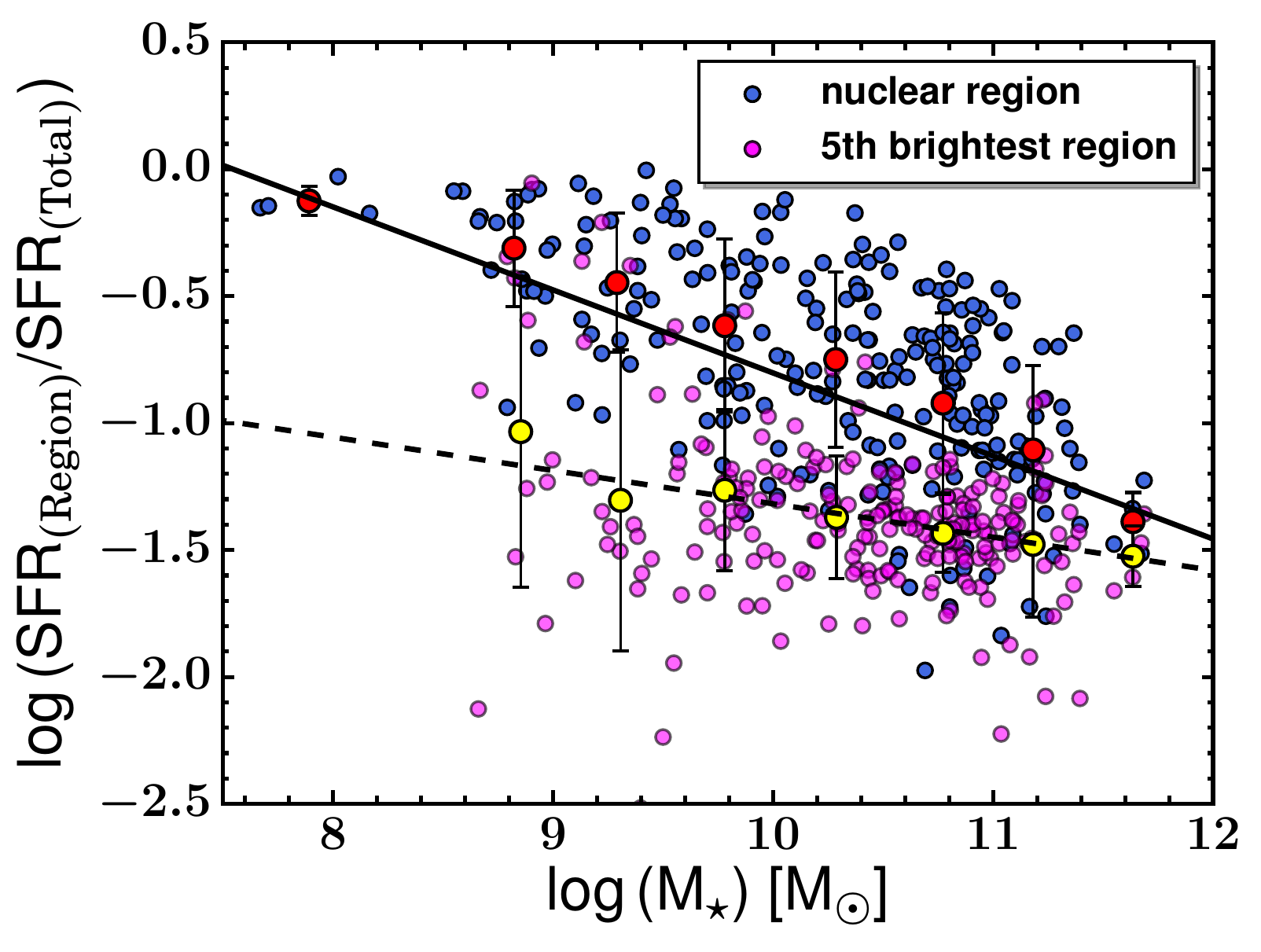}
		\caption{The ratio of nuclear (central 1kpc diameter) to total SFR as a function of total stellar mass (blue points) and the corresponding ratio of the 5th brightest off-nuclear region to total SFR (magenta points). The red points are the nuclear-to-total SFR medians in mass bins and the yellow points are the medians of the 5th brightest region to total SFR, both shown with 1$\sigma$ error bars. Due to the small number of galaxies at the low-mass end, we calculated the mean SFR ratio there in a wider bin, from 7.5--8.5 $\log(\textrm{M}_{\star}/\textrm{M}_\odot)$, but all other medians were calculated in 0.5 dex mass bins.}
		\label{SFR_ratio}
	\end{center}
\end{figure}

The (circum-)nuclear regions of galaxies follow the same correlation as the rest of the sub-galactic regions (Figure \ref{SGMS3panel}, left panel) with similar scatter 0.34 dex. Not only do the nuclear regions follow the normal SGMS, they also closely correlate with total galaxy stellar mass (Figure \ref{MS-NMS}). The scatter is 0.39\,dex, moderately larger than for the MS and SGMS. The nuclear MS (NMS) separates from the global MS with increasing stellar mass. To investigate the relation between the total and nuclear SFR at different masses, we measured the ratio of nuclear to total SFR in the entire stellar mass range as well as in mass bins (Figure \ref{SFR_ratio}) and performed linear regression fitting for the binned values to quantify the ratio. The best-fit line is expressed by: 
\begin{equation} \textrm{log} (\frac{\textrm{SFR}_{\textrm{nuclear}}}{\textrm{SFR}_{\textrm{Total}}}) = (-0.37\pm 0.02)\, \log(\frac{M_{\star}}{10^{10}\textrm{M}_{\odot}}) + (-0.73\pm 0.02) 
\end{equation} 
The scatter around the best-fit line is 0.33 dex. Because the bulge component in massive galaxies is larger relative to the disk, the ratio of nuclear to total SFR decreases with stellar mass. This is in agreement with our result (Figure \ref{SFR_ratio}). In contrast, the disk SFR is expected on average to show less variation with increasing stellar mass. To verify this, we measured the ratio of SFR in the 5th brightest off-nuclear region (which is well separated from the nuclear region) to total SFR with respect to total $ M_{\star} $ using the 230 galaxies having at least 5 apertures. Indeed the average $\rm SFR_{Region}/SFR_{Total}$ (Figure \ref{SFR_ratio}) is nearly independent of galaxy mass, having a slope of $-0.15$ and a scatter of 0.31~dex.

\section{Discussion} \label{Discussion}

Although our derived (SG)MS parameters are within the ranges reported in the literature, there are still differences between our results and studies at similar redshifts. These differences reflect the diversity in the definition and selection of SFGs samples as well as the methods used to determine SFR and stellar mass. 

\subsection{The global MS of star formation}

Previous authors have found that MS slopes range from $\alpha = 0.6$ to 1 (\citealt{Rodighiero11}; see also Table 4 of S14). Low-$ z $ studies, comparable with the redshift range of our sample, e.g., \cite{Whitaker12} (W12) or results from the CALIFA (\citealt{CALIFA}) survey (e.g., D16; C16) reported sub-linear slopes, while \cite{Elbaz11} on the other hand found a MS slope of 1. As discussed in detail by S14, this range of MS slopes is attributed to a number of factors including the methods used to determine SFR and stellar mass. Most of the variation however arises from sample selection. Selecting only actively star-forming galaxies produces MS slopes closer to unity, while a mixture of star forming and passively evolving populations will produce sub-linear slopes (S14). W12 showed that massive quiescent galaxies have lower sSFRs and that the MS slope depends on galaxy colors, with redder dust-attenuated populations having shallower slopes than blue, low-dust galaxies. Furthermore, color-based criteria such as the \textit{BzK} two-color selection or more generally any technique that selects blue objects will emphasize the most active SFG populations, producing steep MS slopes (\citealt{Karim11}). 

Our representative sample of nearby IR-selected star-forming galaxies defines a MS with $ \alpha_{\textrm{MS}} = 0.98$. This result is in perfect agreement with \cite{Elbaz11}, who also used an IR-selected low-$z$ sample similar to the SFRS. On the other hand, our result differs from the sub-linear MS slopes found in other low-$z$ works, mostly because of sample selection effects. For instance, W12 used a two-color diagram ($U-V$ versus $V-J$) to mostly exclude quiescent galaxies from their analysis and found $ \alpha_{\textrm{W12}} = 0.6 $. C16 used the CALIFA survey clipping out high-inclination galaxies and therefore not mapping the entire range of conditions in SFG populations and found $ \alpha_{\textrm{C16}} = 0.81 $.

Unrecognized AGN emission can result in unreliable SFR and stellar-mass determinations, affecting the MS slopes in ways difficult to quantify. W12 used rest-frame colors to distinguish between quiescent and star-forming galaxies, but AGN or composite objects were not necessarily identified or excluded. D16 excluded only type-1 Seyferts from their sample without accounting for obscured or low-luminosity AGNs. Our approach distinguishes between different activity classes and selects pure SFGs with a combination of IR and optical emission-line diagnostics (Maragkoudakis et al. in preparation). In this way we are able to identify and exclude obscured AGNs and galaxies having contribution from both star-formation and AGN activity (composite objects), providing a clean sample of bona-fide SFGs. 

The way stellar masses are determined can lead to additional discrepancies in the MS. Most mass-to-light (\textit{M/L}) ratio conversions assume a galaxy color dependence, the definition of which relies on the assumed model SFH and metallicity. Different combinations of SFH and metallicities will yield different mass-to-light ratio calibrations, resulting in different derived stellar masses between different studies. In addition, for a given color-\textit{M/L} calibration, dust reddening corrections affect galaxy color determination, introducing additional scatter in the derived stellar masses. The above effects can also be important when considering a mass dependence of the MS slope with a curvature form description (e.g., \citealt{Whitaker14}; \citealt{Lee15}, \citealt{Schreiber15}) rather than a linear form. Differences between stellar mass calibrations cause scatter in the horizontal location of a galaxy in the SFR--$M_{\star} $ plane, but assuming a mass-dependent MS slope (i.e., curvature) can also increase the difference between the derived MS slopes reported in the literature. W12 calculated stellar masses using SED modeling, while D16 summed the individual spaxel masses derived from stellar population fits to the optical spectrum. In both cases the results depend on several parameters such as the assumptions of SFHs, the choices of stellar population models, stellar metallicities, dust treatment, or the sampling of these parameter spaces (\citealt{Conroy13}). Moreover, for the galaxies hosting AGNs, the SED modeling and stellar population synthesis methods can greatly overestimate stellar mass when the AGN continuum is not properly treated, i.e., when both the AGN and stellar components are modeled only with stellar populations. The SFRS stellar masses in this paper were determined photometrically using IRAC 3.6 $\micron$ measurements in a manner consistent with the well-calibrated $ K $-band based determinations (Bonfini et al. in preparation). Because these \textit{M/L} calibrations are less dependent on stellar population details and dust reddening, they produce robust stellar mass estimations.

Different SFR calibrations can also influence the derived MS properties. Luminosity-to-SFR conversions can vary by up to $\sim$50\% and create large offsets between different studies (e.g., KE12, S14). \cite{Davies16} showed that even when the most commonly used luminosity to SFR calibrations are re-defined and re-calibrated to each other, there is still a non-linear relation between them, and therefore different SFR indicators will produce different SFR--$M_{\star} $ slopes. W12 determined SFRs from the combined UV and total IR emission. The total IR luminosity was estimated from a conversion based only on the observed 24\,\micron\ flux density and was derived from a single template that is the log average of the \cite{DH02} templates. The 24\,\micron\ flux however is sensitive to AGN emission, so for AGN hosts some fraction of 24 $ \micron $ emission must originate in those AGN. D16 used the stellar population synthesis results of the mass initially turned into stars, with an arbitrary choice of star formation time scale for the determination of SFR. These results are model-dependent and especially unreliable for galaxies hosting AGNs. The use of IRAC 8.0 $\micron$ avoids these difficulties and provides sufficient spatial resolution to isolate distinct regions within galaxies. 

\subsection{The Sub-galactic Main Sequence}
The SGMS holds at $\sim$1 kpc scales with a slope of $ \alpha_{\textrm{SGMS}} = 0.91 $ and dispersion of $ \sigma_{\textrm{SGMS}} = $ 0.31 dex, similar to the integrated MS. This similarity implies that the on-going star formation and stellar mass buildup from past star formation trace each other on both global and sub-galactic scales. However, because star formation is a local process, its sub-galactic manifestation and scaling relations are probably more fundamental compared to their integrated analog. This view is supported by the SGMS from randomly drawn sub-galactic regions (Figure \ref{SGMS-morph}). Random regions produce the same SGMS slope and little more scatter than for all regions combined, implying that any individual region of any galaxy is likely to be representative of the same $\Sigma_{\textrm{SFR}}$--$\Sigma_{\star}$ correlation as the global SGMS. 

Our derived SGMS slope is consistent with W13 and \cite{Magdis16}, who reported $ \alpha_{\textrm{W13}} = 0.95 $ and $\alpha_{\textrm{M16}} = 0.91 $ respectively while sampling similar or larger physical scales. Nevertheless, while C16 have presented sub-galactic MS results in 0.5 -- 1.5 kpc scales showing a shallower slope ($\alpha_{\textrm{C16}} = 0.68 $), extra caution is needed when the global SFR calibrations are used on scales smaller than 1 kpc. Below 1 kpc several factors introduce stochasticity. Among the most important are incomplete sampling of the IMF, leading to large fluctuations in the luminosity of the SFR tracers, the non-applicability of the continuous star formation assumption embedded in the global SFR recipes, or in cases where the regions are smaller than the Str\"{o}mgren diameters of H\textsc{ii} regions, indirect tracers such as H$ \alpha $ can trace diffuse ionized gas located several parsecs away from any young star (see KE12 for details). These effects would introduce scatter to the derived SFR, possibly resulting in the sub-linearity reported by C16 or D16.

The SGMS slope increases from Sa to later types. The ``late-type" galaxies (Sc--Irr) show a linear $\Sigma_{\textrm{SFR}} - \Sigma_{\star}$ relation ($\alpha = 0.97$), compared to the sub-linear slope ($ \alpha = 0.81 $) of the Sa--Sbc group (Figure \ref{SGMS-morph}), bearing the closest resemblance to the global MS characteristics. The sub-linearity in this group indicates that the sSFR in the most massive regions is not as high as in the disk. In contrast, the unity slope in ``late-types" suggests that the sSFR in all sub-galactic regions is similar regardless of local mass, and all regions of the galaxy grow at a similar rate. On the other hand, D16 found that while the global MS slope increases from Sa to Sd Hubble types, the corresponding resolved MS slopes are fairly similar.

The morphology dependence of the SGMS slope can be attributed to the way in which gas is distributed in different galaxy types. \cite{Magdis12} showed that gas fractions correlate strongly with sSFR, i.e., that $ M_{gas}/M_{\star} $ increases strongly with sSFR at a given redshift. \cite{Tacconi13} also showed that at constant stellar mass, a galaxy's SFR is mainly driven by the available molecular gas reservoir. Furthermore, the non-linear relationship between the disk-averaged surface densities of star formation and total (atomic and molecular) gas, described by the Kennicutt-Schmidt (KS) law, has a slope of $\sim$1.4--1.5 (KE12) and states that $\Sigma_{\textrm{SFR}}$ is higher in regions of higher gas density compared to lower gas density regions. Combining the SGMS description ($\Sigma_{\textrm{SFR}} \propto \Sigma_{\star}^{a} $) with the KS law ($\Sigma_{\textrm{SFR}} \propto \Sigma_{gas}^{1.4}$), gas mass density scales with stellar mass density as $ \Sigma_{gas} \propto \Sigma_{\star}^{\alpha/1.4}$. This yields that gas concentration, in general, is higher in more massive sub-galactic regions, and this effect is stronger in ``late-type" galaxies with an SGMS slope $\alpha = 0.97$ ($ \Sigma_{gas} \propto \Sigma_{\star}^{0.69}$) compared to ``early-type spirals" with slope $ \alpha = 0.81 $ ($ \Sigma_{gas} \propto \Sigma_{\star}^{0.58}$).

The ``lenticulars" (E--S0/a) group has a steep SGMS slope ($\alpha = 1.09$; Figure \ref{SGMS-morph}). This implies that while overall star forming activity in early types is considerably lower due to limited gas, the regions within them that are forming stars have similar properties with later Hubble types. The small number of early-type galaxies (E--S0) in our sample may on one hand not be representative of the general early-type populations, possibly sampling those with highest activity, but nevertheless our current results shows that star formation in bulge-dominated systems follows a similar trend as those in later Hubble types.

The SGMS of all morphologically classified galaxies combined (Figure \ref{SGMS_All_Morph}) has the same properties as the ``all-galaxies" SGMS (of known and unknown morphological Hubble types) showing the same slope and dispersion. In contrast, the morphologically unclassified galaxies have a shallower slope of $ \alpha = 0.72 $ (Figure \ref{SGMS_All_Morph}). The main cause for this difference is that the bulk of unclassified galaxies occupy only the mid to high-stellar mass end with only 8\% below $\log(\textrm{M}_{\star}/\textrm{M}_{\odot}) = 9.5$. As shown by \cite{Lee15}, galaxies more massive than $ \textrm{M}_{\star} \gtrsim 10^{10}$ have a much lower average sSFR, and  additionally the MS appears to flatten at masses above that characteristic mass. Star formation in low-intensity outer disk regions is generally more stochastic and bursty, and the existence of extended UV disks at extreme galactocentric distances (e.g., \citealt{Thilker05}; \citealt{GildePaz05}) supports this view. \cite{Thilker07} showed that certain extended UV disk galaxies show UV-bright/optically faint emission features beyond the anticipated location of the star formation threshold (e.g., \citealt{Martin01}), which can therefore explain the increasing SGMS slope in regions of lower surface brightness. In addition, the ``early-type spirals" and ``late-types" groups have a fairly good sampling of the lower-stellar mass end giving a more thorough description of the SGMS. Out of the morphologically classified galaxies, the ``late-types" being both the most numerous in our sample and having the largest number of apertures relative to the other morphological groups, appear as the main drivers of both the MS and SGMS. This is further justified from the fact that the ``late-types" SGMS has closer characteristics (slope and normalization) to the ``all-galaxies" SGMS.

Single-galaxy SGMSs formed from regions within individual galaxies are similar to the SGMS formed from all regions in all galaxies. The average sSFR measured from the local SFR/$M_{\star} $ ratio of all sub-galactic regions within each galaxy is within 0.3 dex of the total galaxy sSFR (Figure \ref{isSFR_vs_sSFR}), further supporting the view that kpc-scale star-forming activity is representative of the total. Given the power-law form of the SGMS (SFR $ \propto M^{\alpha}$), the sSFR is accordingly described as: SFR/$M_{\star} $ $ \propto M^{\alpha -1}$. Therefore, galaxies with steeper individual SGMS slopes lie above the line of equality in Figure \ref{isSFR_vs_sSFR} while those with shallower slopes lie below. In addition, bulge-dominated systems are expected to have shallower average sSFR measured from their sub-galactic regions, because the sSFR in the massive bulge regions is generally lower compared to the disk. These systems would lie below the line of equality in Figure \ref{isSFR_vs_sSFR}. The morphologically unclassified galaxies are predominately located bellow this line, indicating massive bulge-dominated systems. This is in agreement with the fact that the unclassified group, constituting mostly massive galaxies, has a shallower SGMS slope (0.72). The average SGMS slope of individual galaxies ($ \alpha = 0.95 $) coincides with the ``all-galaxies" SGMS slope ($ \alpha_{\textrm{MS}} = 0.91$). Galaxies with total $M _{\star} $ $ > 10^{10} \,\textrm{M}_{\odot}$ or total SFR $ > 6\, \textrm{M}_{\odot}/\textrm{yr}$ have sub-linear ($ \alpha < 1 $) SGMS slopes, meaning that the sSFR declines with mass. This is expected in earlier Hubble types, and it is evident in the ``early-type spirals" morphological group which contains most of the high-$M _{\star} $ and high-SFR end. This result also underlines that incomplete samples or samples selected at mid to high SFR or stellar mass will report a sub-linear SGMS, similar to the results of the morphologically unclassified galaxies in our sample. Our analysis, in a complete sample covering 4 orders of magnitude in both total SFR and $M _{\star} $,  shows that the SGMS is not necessarily sub-linear. In contrast the ''late-types" group is mostly found at the low-$M _{\star} $ low-SFR end with super-linear slopes. Therefore, a complete framing of the SGMS requires both a wide range in the SFR--$M _{\star} $ space and a mixture of morphological types.

\subsection{The nuclear and circumnuclear regions}
Despite the spatially resolved correlation observed for star-forming regions in galaxies, the nuclear regions have not been investigated yet in the literature. Our study shows that the nucleus is among the most active (if not the most active) regions of star formation (Figure \ref{SFR_hist}), even in the case of the gas-rich late-type galaxies where star formation mainly occurs in the midplane of the thin disks. Similarly, the nucleus has the highest stellar mass concentration (Figure \ref{Mass_hist}), as expected, because it resides at the center of bulge. The nuclear $\Sigma_{\textrm{SFR}} - \Sigma_{\star}$ correlation, formed exclusively from the nuclear regions of SFGs \footnote{Because the nuclear region potentially encompasses part of the circumnuclear star-forming activity, the nuclear SGMS should be better considered as a ``circumnuclear" sequence.}, is almost as tight as the SGMS and MS.  
 
The existence of the NMS suggests a connection between the global and (circum-) nuclear SFR. At the lowest stellar-mass end, the two sequences overlap, which is expected because this end is mostly populated by dwarf galaxies with major axes of $\sim$4--4.5 kpc, only slightly larger than the aperture used in the analysis. This effect diminishes as we examine more massive and larger galaxies, where the two sequences become distinct. The average logarithmic ratio of nuclear to total SFR (Figure \ref{SFR_ratio}) in the SFRS sample is $\log(\textrm{SFR}_{\textrm{(nuclear)}}/\textrm{SFR}_{\textrm{(total)}}) = -0.77$, but the value strongly depends on galaxy mass. The NMS can potentially be a key ingredient in understanding star-forming activity in the nuclear region and its connection to star formation in the main body of the galaxy. For instance, stars that are forming now in the centers of galaxies are the building blocks of bulges. Similarly, past (higher-$z$) star-forming activity in the nuclear regions of galaxies are responsible for the formation and the growth of present day (lower-$z$) bulges. The NMS can therefore provide evolutionary insights on the secular growth of bulges with respect to the total stellar masses of galaxies and their regulation through AGN. Because AGN accretion drives black hole growth, the NMS and its relation to the (SG)MS can be an alternative tool to the average black hole growth and mean stellar mass relation reported by \cite{Mullaney12} to study growth of black holes and their influence on the nuclear/circumnuclear star formation.

The NMS and its relation to the total SFR may also be an asset in the study of AGN populations and specifically in the still open subject of AGN feedback. For instance, SED modeling codes (e.g., CIGALE; \citealt{CIGALE1}, \citealt{CIGALE2}) can simultaneously model and measure the AGN and star-forming luminosity components of galaxies. Comparing the nuclear SFR for a non-AGN host at a given stellar mass via the NMS and the corresponding nuclear SFR of an AGN host at the same stellar mass through SED fitting can provide insights on AGN feedback. Lower SED-derived SFRs compared to NMS-derived SFRs would be indicative of AGN-driven quenching of star formation (negative feedback), while the opposite case would imply AGN-driven boosting of star formation (positive feedback).

\section{Conclusions}
Based on a representative sample of local galaxies with a broad range in stellar mass ($7.7 < \log(M_{\star}/\textrm{M}_{\odot}) < 11.7$) and SFR ($ -2.2 < \log(\textrm{SFR}/\textrm{M}_{\odot}\textrm{yr}^{-1}) < 2.1 $), we have examined the MS of star formation in global and sub-galactic scales. Our main conclusions are as follows.
\medskip

(i) The global SFR--$M_{\star} $ correlation of SFGs at low redshifts has a linear form with a slope of 0.98. The wide range of four orders of magnitude in stellar mass and SFR sets strong constraints on the derived slope and normalization of the MS, and combined with the fact that the SFRS sample is fully representative of the star-forming conditions in the local Universe, this work provides a comprehensive description of the MS of SFGs. 

(ii) The correlation between $\Sigma_{\textrm{SFR}}$ and $\Sigma_{\star}$ at kpc spatial scales within galaxies described by the SGMS has similar characteristics to the global MS. Because star-formation is primarily a small-scale phenomenon, the SGMS can be considered as more fundamental. This assertion is supported by the SGMS produced from random regions within galaxies being on average representative of the global MS characteristics.

(iii) The SGMS produced from sub-galactic regions of individual galaxies has on average the same characteristics as the composite SGMS from all galaxies. Sub- or super-linearity in the SGMS slope may be reported when the observed SFR--$M_{\star} $ range is limited or samples are not fully representative.

(iv) The slope of the SGMS differs among different Hubble types, increasing from Sa to Irr, explained by the variations of gas content between different morphological types. Furthermore, while bulge-dominated populations are considered quiescent, our S0--S0/a group shows that when star-forming activity is present it has similar characteristics to those in later Hubble types.

(v) The circumnuclear $\Sigma_{\textrm{SFR}}$ has a similarly tight correlation with the $\Sigma_{\star}$ compared to the rest sub-galactic regions.

(vi) The nuclear SFR correlates with total stellar mass forming a nuclear MS, which can be used to examine the bulge and BH growth of galaxies at different redshifts as well as characterize feedback processes (positive or negative) between AGN and star-forming activity. The ratio of nuclear to total SFR is mass-dependent, consistent with a slope of $-0.36$.

In future work we intend to investigate the dependence of the SGMS parameters on the size of the physical regions examined by sampling varying sized regions and quantifying the impact of aperture effects. Our ongoing H$ \alpha $ imaging campaign at the 1.3-m Ritchey-Cr\'{e}tien telescope at the Skinakas Observatory in Crete will allow us to derive H$ \alpha $-based SFRs for the same sample of galaxies, addressing the dependence of the SGMS on different SFR calibrations. Lastly, using SED modeling, we will be able to isolate the AGN contribution in AGN hosts and deliver the SGMS of different AGN populations as well as estimate the total and resolved sSFR functions of galaxies along the (SG)MS plane in different activity types.

\section*{ACKNOWLEDGMENTS}

We would like to thank the referee for the constructive comments and suggestions which have improved the clarity of this paper. AM and AZ acknowledge funding from the European Research Council under the European Union’s Seventh Framework Programme (FP/2007-2013) / ERC Grant Agreement n. 617001. This work is based [in part] on observations made with the Spitzer Space Telescope, which is operated by the Jet Propulsion Laboratory, California Institute of Technology under a contract with NASA. For this research, we have made extensive use of the Tool for Operations on Catalogues And Tables (TOPCAT) software package (\citealt{TOPCAT}).

\bibliographystyle{mnras}
\bibliography{Bibliography}

\end{document}